\documentclass[12pt]{article}
\usepackage{amsmath} 
\usepackage{amssymb} 
\usepackage{graphics} 
\usepackage{cite}
\usepackage{epsfig}
\usepackage{color}

\newcommand  {\EL}       {{\it Europhys.\ Lett.\ }}

\newcommand  {\Phy}      {{\it Physica\ }}

\newcommand  {\PNAS}     {{\it Proc.\ Natl.\ Acad.\ Sci.\ USA\ }}

\newcommand{\beq}{\begin{equation}}
\newcommand{\eeq}{\end{equation}}
\newcommand{\beqa}{\begin{eqnarray}}
\newcommand{\eeqa}{\end{eqnarray}}
\newcommand{\bea}{\begin{eqnarray}}
\newcommand{\eea}{\end{eqnarray}}
\newcommand{\dst}{\displaystyle}

\newcommand{\TRUE} {\textrm{true}}

\newcommand{\T}    {\textrm{T}}
\newcommand{\F}    {\textrm{F}}

\newcommand{\Odef} {O_{\textrm{\scriptsize default}}}
\newcommand{\weq}[1][N]{w^{#1}_{\textrm{\scriptsize eq}}}
\newcommand{\Q}    {\mathbf{Q}}
\newcommand{\QNL}  {{\cal Q}_L^N}
\newcommand{\PNL}  {{\cal P}_L^N}
\newcommand{\nb}   {\mathbf{n}}

\newcommand{\nbh}  {\hat{\mathbf{n}}}
\newcommand{\xb}   {\mathbf{x}}

\newcommand{\x}    {\mathbf{x}}
\newcommand{\A}    {\mathbf{A}}

\newcommand{\phiC} {\phi_L^\textrm{\scriptsize C}}
\newcommand{\phiI} {\phi_L^\textrm{\scriptsize I}}
\newcommand{\ip}   {{\acute{\imath}}}

\newcommand{\CL}[1][L]{\langle C_{#1}\rangle_N}
\newcommand{\OL}[1][L]{\langle\Omega_{#1}\rangle_N}
\newcommand{\CLinf}[1][L]{\langle C_{#1}\rangle_\infty}
\newcommand{\OLinf}[1][L]{\langle\Omega_{#1}\rangle_\infty}
\newcommand{\rC} {r^\textrm{\scriptsize C}}
\newcommand{\rI} {r^\textrm{\scriptsize I}}
\newcommand{\ri} {r_\infty}
\newcommand{\dr} {\Delta r}
\newcommand{\dri}{\Delta r_\infty}
\newcommand{\nat}{\mathbb{N}}
\newcommand{\trm}{\textrm}
\newcommand{\srm}[1]{\textrm{\scriptsize{#1}}}

\begin{document}

\begin{flushright}
LU TP 04-10\\
May 10, 2004
\end{flushright}

\vspace{0.2in}

\begin{center}

\large
{\bf Genetic Networks with Canalyzing Boolean Rules}\\
{\bf are Always Stable}

\vspace{0.2in}
\normalsize

{\bf Stuart Kauffman$^{1*}$, Carsten Peterson$^{2*}$,}\\ 
{\bf Bj\"orn Samuelsson$^{2*}$ and Carl Troein$^{2*}$}


\normalsize 

$^1$Department of Cell Biology and Physiology, \\ 
University of New Mexico Health Sciences Center, Albuquerque, NM 87131, USA

$^2$Complex Systems Division, Department of Theoretical Physics\\
Lund University,  S\"{o}lvegatan 14A,  S-223 62 Lund, Sweden \\
{\tt http://www.thep.lu.se/complex/}

$^*$All authors have contributed equally
%

Submitted to {\it Proceedings of the National Academy of Sciences USA}

\end{center}

\vspace{0.60in}

\normalsize

\noindent
{\it Corresponding author:} Carsten Peterson; carsten@thep.lu.se

\noindent
{\it Subject Category:} Biological Sciences: Biophysics.


\vspace{0.1in}
\noindent
{\it Manuscript information:} 19 text pages, 4 figures and 0 tables; 
184 words in abstract; 46980 total characters in the paper; supporting 
text for the web.

\newpage

\begin{center}
{\large \bf Abstract}
\end{center}

\noindent


We determine stability and attractor properties of random Boolean genetic 
network models with canalyzing rules for a variety of architectures. For all 
power law and exponential as well as flat in-degree distributions, we find that 
the networks are dynamically stable. 
Furthermore, for architectures with few inputs per node, the dynamics 
of the networks is close to critical.
In addition, the fraction of genes that are active decreases with 
the number of inputs per node.
These results are based upon investigating ensembles of networks using 
analytical methods.
Also, for different in-degree distributions, the numbers of fixed points
and cycles are calculated, with results intuitively consistent with the
stability analysis; fewer inputs per node implies more cycles, and vice versa. 
There are hints that genetic networks acquire broader degree
distributions with evolution and hence our results then indicate that for single
cells, the dynamics should become more stable with evolution. However, such
an effect is very likely compensated for by multicellular dynamics,
since one expects less stability when interactions between cells are included.
We verify this by simulations of a simple model for interactions between cells.    
 
\newpage

\section{Introduction}

With the advent of high-throughput genomic measurement methods, it is soon
within reach to apply reverse engineering techniques and map out genetic 
networks inside cells. 
These networks should perform a task, and, very importantly, be stable 
from a dynamical point of view.
It is therefore of utmost interest to investigate the generic properties of 
networks models, such as architecture, dynamic stability and degree of 
activation as functions of system size. 
Random Boolean networks have for several decades received much attention in 
these contexts. These networks consist of nodes, 
representing genes and proteins, connected by directed edges, representing
gene regulation.
The number of inputs to and outputs from each node, the in- and out-degrees, 
are drawn from some distribution.

It has been shown that with a fixed number, $K$, of inputs per 
node such network models exhibit some interesting properties\cite{kauffman}.
Specifically, for $K=2$ the dynamics is critical, {\it i.e.}, right
between stable and chaotic.
Furthermore, one might interpret the solutions, fixed points and 
cycles, as different cell types. Being critical is considered advantageous 
since it should promote evolution. These results were obtained with no 
constraints on the architectures and assumed a flat distribution of the 
Boolean rules.

It appears natural to revisit the study of Boolean network ensembles,
given recent experimental hints on network architectures and rule 
distributions. For transcriptional networks,
there are indications from extracted gene--gene networks that, for
both {\it E. coli} \cite{ecoli} and yeast, \cite{young} the out-degree
distribution is of power law nature. The in-degree distribution appears to
be exponential in {\it E. coli}, whereas it may equally well be a
power law in yeast. In \cite{aldana} stability properties of 
random Boolean networks were probed with power law in-degree distributions,
and regions of robustness were identified. 

The distribution of rules is likely to be structured and not random. 
Indeed, in a recent compilation \cite{harris} (see also \cite{yeast}), all 
rules are canalyzing \cite{kauffman}; a canalyzing Boolean function has at least 
one input, such that for at least one input value, the output value is fixed. 
It is not straightforward to generate biologically relevant canalyzing 
functions. In \cite{yeast} the notion of nested canalyzing functions was 
introduced, which facilitates generation of rule distributions.

We find that networks with nested canalyzing rules are stable,
for {\it all} power law and exponential in-degree distributions. Furthermore, as 
the degree distribution gets flatter, one moves further away from criticality. 
Also, the average number of active genes (fraction of genes that take the
value `true')  is predicted for different powers.

There are experimental hints that the in-degree distributions get 
flatter with genome size. This could be understood intuitively, since higher
organisms in general have acquired more complexity in terms of redundancy 
in signal integration.
In such a picture, our robustness analysis indicates that with multicellular
species one would move away from critical dynamics, thereby making evolution less 
accessible. However, the picture is complicated by the presence of extracellular 
interactions. We model these with a simple scheme allowing for 5--10\% 
extracellular traffic, and, not unexpectedly, the system, though still 
robust, moves towards criticality.

\section{Methods and Models}

\subsection{Degree Distributions}

Our results turn out to be qualitatively equivalent for power law, 
exponential and flat in-degree distributions. (By flat we mean a uniform
distribution for up to $K_{\textrm{\scriptsize max}}$ inputs.) In what follows, we choose
to illustrate the results with a power law 
(often denoted {\it scale-free}) distribution, with a cutoff in the
number of inputs, $K$,
\beq
  p^N_K \equiv P(\textrm{\#inputs} = K)
  \propto \left\lbrace
    \begin{aligned}
         \frac{1}{K^{\gamma}}& &&\trm{if $1\leq K\leq N$}\\
         0\,\,\,& &&\trm{otherwise}
    \end{aligned}\right.~.
\label{p_in}
\eeq
where $N$ is the number of nodes.
In yeast protein--protein networks \cite{scalefree}, and also in other 
applications, {\it e.g.}, the Internet and social networks,
$\gamma$ appears to lie in the range 2 to 3. In our calculations, we
explore the region 0 to 5, varying  $N$ from 20 to infinity.
The connectivity of gene--gene 
networks extracted from yeast \cite{young} appear to behave like in 
Eq.~\ref{p_in} for the in-degree distributions, with an exponent $\gamma$ in the 
range 1.5--2. For {\it E.~coli} \cite{ecoli} data, an exponential fits 
somewhat better than a power law, but the data are statistically inconclusive.
For the mammalian cell cycle genes, slightly lower $\gamma$ has been
extracted \cite{hill}.
The average number of inputs varies with $N$ and $\gamma$, and grows with
decreasing $\gamma$. For very high $\gamma$, it is
essentially 1. In Fig.~1, typical network realizations for $N=20$ are shown
for $\gamma=1$, 2 and 3, respectively.

\subsection{Canalyzing Boolean Rule Distributions}

In most studies of Boolean models of genetic networks, all Boolean 
rules have been employed \cite{kauffman}. In a previous paper, we 
introduced {\it nested canalyzing rules} and showed that it is possible 
to generate a distribution of such rules, that fits well with rule
data from the literature \cite{yeast}.

A {\it canalyzing rule} is a Boolean rule with the property that
one of its inputs alone can determine the output value, for either
`true' or `false' input. This input value is referred to as
the {\it canalyzing value}, the output value is the {\it canalyzed value},
and we refer to the rule as being {\it canalyzing on} this particular input.

Nested canalyzing rules are a very natural subset of canalyzing rules,
stemming from the question of what happens in the non-canalyzing case.
That is, when the rule does not get the canalyzing value as input,
but instead has to consult its other inputs. If the rule then is
canalyzing on one of the remaining inputs, and again for that input's
non-canalyzing value, and so on for all inputs, we call the rule nested
canalyzing. All but six of the roughly 150 observed rules were nested
canalyzing \cite{yeast}.

With $I_m$ and $O_m$ denoting the canalyzing and canalyzed values,
respectively, and suitable renumbering of the inputs, $i_1, \ldots, i_K$,
the output, $o$, of a nested canalyzing rule can be expressed
on the form
\beq
  o = \left\lbrace\begin{array}{ll}
    \!\!O_1 & \!\trm{if } i_1 = I_1 \\
    \!\!O_2 & \!\trm{if } i_1 \neq I_1\,\trm{and}~i_2 = I_2 \\
    \!\!O_3 & \!\trm{if } i_1 \neq I_1\,\trm{and}~i_2 \neq I_2
           \,\trm{and}~ i_3 = I_3 \\
     \vdots   & \\
    \!\!O_K & \!\trm{if } i_1 \neq I_1 \,\trm{and}\cdots
                      \trm{and}~ i_{K-1} \neq I_{K-1}
                      \,\trm{and}~ i_K = I_K \\
    \!\!\Odef & \!\trm{if } i_1 \neq I_1 \,\trm{and}\cdots
         \trm{and}~ i_K \neq I_K~. \\
  \end{array} \right.
\eeq

For each value of $K$, we generate a distribution of $K$-input rules,
with the inputs to each rule taken from distinct nodes. All
rules should depend on every input, and this dependency
requirement is fulfilled if and only if $\Odef = \trm{not}\,O_K$.
Then, it remains to choose values for $I_1, \ldots, I_K$ and
$O_1, \ldots, O_K$. 
These values are independently and randomly drawn with the probabilities
\beq
  P(I_m = \TRUE) = P(O_m = \TRUE) = \frac{\exp(-2^{-m}\alpha)}
         {1 + \exp(-2^{-m}\alpha)}
\label{eq: nestcan}
\eeq
for $m = 1,\ldots,K$, where $\alpha$ is a constant.
Eq.~\ref{eq: nestcan} can be seen
as a way to put a penalty both on outputting `true' and on letting
a `true' input determine the output. More precisely: Let $f$ be the
fraction of `true' outputs in the truth table, and let $g$ be the
fraction of input states such that the first input that has its
canalyzing value is `true'. Then, the probability to retrieve a
specific rule is proportional to $\exp[-\alpha(f+g)/2]$.

Our rule distribution fits observed data well \cite{yeast}, given that
$\alpha = 7$. For all generated distributions, we keep $\alpha = 7$.
A high value of $\alpha$ means a high penalty on active genes,
while $\alpha = 0$ means equal probabilities for activity and inactivity.

\subsection{Robustness Calculations}

We wish to address the question of robustness in network models.  In a
stable system, small initial perturbations should not grow in time.
In \cite{yeast}, this was probed by monitoring how the Hamming distance
$H$ between random initial states evolved in a ``Derrida plot''
\cite{derrida}. Specifically, the slope in the low-$H$ region shows
the fate of a small perturbation after a single time step.  This
implicitly assumes that `true' and `false' are equally
probable in the initial states.
 
Our chosen distribution of Boolean rules will favor `false' node
values.  Preferably, a robustness measure should reflect the network
properties in the vicinity of the equilibrium distribution, 
where the in- and out-degree distributions of `true' and `false'
are identical. See {\it Appendix A}. We therefore define the
robustness $r_N$ for an ensemble of $N$-node networks as the average
effect, after a single time step, of a small perturbation at
this equilibrium distribution.

To compute $r_N$, we introduce the total sensitivity, $S(R)$,
of a given $K$-input rule $R$. $S(R)$ is the sum of the probabilities
that a single flipped input will alter the output of $R$. Thus,
\beq
\begin{split}
  S(R) = \sum_{j=1}^K P\bigl[&R(i_1,\ldots,i_{j-1},0,i_{j+1},\ldots, i_K)\\
         &\ne R(i_1,\ldots,i_{j-1},1,i_{j+1},\ldots, i_K)\bigr]~,
\end{split}
\eeq
where the probability is calculated over the
equilibrium distribution of input values, $i_1,\ldots,i_K$. Then, $r$ is
given by $r = \langle S(R)\rangle$ where the average is taken over
all rules in a given network (see {\it Appendix A}). This also allows us
to calculate $r$ when the rules are drawn from a distribution.
Note that $r$ is calculated without any assumption on how the
inputs to the rules are chosen. This means that $r$ stays the same for
any output connectivity, and for any way to build a network containing
a certain set of rules. In other words, $r$ is a strictly local
stability measure that is independent of whether the network is
divided into some kind of clusters or not.

Let $S_K$ denote the average of $S(R)$ over a distribution of $K$-input
rules. The average robustness of a randomly chosen
network with $N$ nodes is then given by
\beq
  r_N = \sum_{K=1}^N p^N_K S_K~.
\eeq
See {\it Appendix A} on calculation of $S_K$ for nested
canalyzing rules. With the nested canalyzing rule distribution,
defined by Eq.~\ref{eq: nestcan}, $S_K<1$ for $K=2,3,\ldots$, provided
that $\alpha\ne0$. $S_1$ is always 1, since every rule has to depend
on all of its inputs. ($\alpha=0$ yields $S_K=1$ for all $K$.)  This
means that every network ensemble with the canalyzing rule
distribution, and $\alpha\ne0$, that not solely consists of one-input
nodes, is stable in the sense that $r<1$.

\subsection{Number of Attractors}

Attractors in the Boolean model can be seen as distinct cell types
\cite{kauffman}. It is, however, not straightforward to tell which attractors 
are biologically relevant. First, one can ask what cycle lengths are
relevant.  Second, the attractor basin sizes vary in a very broad
range, and attractors with small attractor basins may be biologically
irrelevant.

The broad distribution of attractor basin sizes also means that the
number of attractors found by random sampling is strongly dependent on
the number of samples \cite{prl}. We choose to monitor the number of
attractors, $\CL$, of different periods, $L$, using exact methods
(for the limit $N \rightarrow \infty$)
and full enumerations of the state space (for small networks).

Given that $r < 1$, which means that the network is
subcritical, the average number of attractors of a certain length,
$L$, will approach a constant, $\CLinf$, as the
system size, $N$, approaches infinity. We find an analytic
expression for $\CLinf$, and find that it is qualitatively consistent
with results from a full state space search for $N=20$.

To investigate the limit $N \rightarrow \infty$, we split up the
robustness measure, $r$, into $\rC$ and $\rI$, where $r$
is the average number
of outputs that, after one time step, will be affected by one
flipped node. We define $\rC$ and $\rI$ as the numbers of nodes that
copy respectively invert the state of the flipped node. This means,
{\it e.g.}, that $r = \rC + \rI$. For convenience, we define $\dr = \rC -
\rI$.

$\CLinf$ can be expressed as a function of
$r_\infty$ and $\dr_\infty$ for each $L$. For $L=1,2,3$ we get
\begin{align}
  \CLinf[1] &= \frac{1}{1-\dri}\\
  \CLinf[2] &= \frac{\ri-\dri+\ri\dri}{2(1-\ri)(1-[\dri]^2)}\\
  \CLinf[3] &= \frac{\ri^3+(\dri)^3-\ri^3(\dri)^3}{3(1-\ri^3)
                (1-\dri)(1-[\dri]^3)}~.
\end{align}
See {\it Appendix B}, where the derivations needed to calculate
$\CLinf$ are presented.

The canalyzing rule distribution
satisfies $\rC > \rI$, meaning that $\dr > 0$. This condition yields
an increased number of fixed points and attractors of odd length,
compared to the symmetric case $\dr = 0$.

It is interesting to note that the limit of the total number of attractors
\beq
  \langle C\rangle_\infty = \sum_{L=1}^\infty\CLinf
\eeq
is convergent for $r<1/2$ and divergent for $r>1/2$. This transition
occurs at $\gamma = 1.376$, with convergence below this value.
See {\it Supporting Text} for details.

\subsection{Tissue Simulations}

In multicellular organisms, we expect communication between cells to
influence the network dynamics. In the real world, there exist several
different types of intercellular signaling. Here we make an initial
exploration using a simple model. Nevertheless, we think the
results reflect some core properties. 

In our model, each cell communicates with its four nearest neighbors 
on a square lattice with periodic boundary conditions. All cells have the 
same genotype, and hence identical internal network architecture and rules. 
Each connection in the network represents how a gene product 
influences the transcription of some gene. Some molecules or signals
can cross cell membranes, and possibly diffuse far, but we only consider
the case of local cell-to-cell signaling at the level of specific
gene--gene interactions.
%
Specifically, a fraction, $\kappa$, of the connections are
flagged as being intercellular, and for such connections the
value is `true' if any of the four neighbors has `true'.

For the overall robustness of such 
tissue networks, it is not sufficient to measure the robustness $r$,
since $r$ only depends on interactions during a single time step,
during which a perturbation only can propagate to the nearest
neighbor cells.
Hence, we desire a multi-step robustness measure, which requires
simulations, since it is outside the scope of our analysis.
Rather than following trajectories from random initial states, we have
chosen to identify fixed points, perturb these randomly by
Hamming distance $H(0)=1$, and then track the mean of $H(t)$
for 20 time steps.
In our simulations, we generate ensembles of networks using  
$5 \times 5$ lattices of cells, where each cell 
contains an identical network of $N=50$ genes. 

\section{Results and Discussion}

Three major findings emerge from Fig.~2, where the average robustness 
$r$ and the fraction of active genes are shown 
as functions of $\gamma$ in Eq.~\ref{p_in}.
\begin{enumerate}
\item The dynamics of the networks is always stable, regardless of
the power in the in-degree distribution. 
\item The stability of the networks increases with the average number of
inputs to the nodes. For flatter distributions,
$r$ approaches the critical value, $r=1$.
\item The average number of active genes decreases with increasing 
in-degree.
\end{enumerate}

Not unexpectedly, the number of attractors increases as the
networks approach criticality, see Fig.~3. This increase is
particularly rapid for long cycles. These
results were obtained with analytical calculation and exhaustive
enumeration of state space. Given the undersampling problems when
simulating Boolean networks \cite{prl}, the feasibility of analytical
calculations is crucial for drawing the firm conclusions above.
However, we did not attempt to extract the distribution of attractor
basin sizes. In future research, it could be interesting to
compare with the exact results for one-input networks in \cite{flyvbjerg}.

The results in Figs.~2 and 3 are shown for power law in-degree distributions. 
However, they are quite general. The corresponding curves for other 
distributions exhibit very similar behaviour, when the $x$-axis ($\gamma$) 
has been transformed to appropriate parameters for other distributions. 
%
In all tested degree distributions, constant nodes ($K=0$) are
excluded. Recall that we also exclude rules that have redundant
inputs. Thus, for low values of the average $K$, most of the rules
will be copy- or invert-operators, which puts the network close to
criticality. The strong stability for wide in-degree distributions,
however, is a result of the canalyzing property of the rules, which
makes forcing structures \cite{kauffman} prevalent.

From analysis of network data from yeast \cite{young}  
and the mammal cell cycle \cite{hill}, there are indications that 
$\gamma$ decreases with the number of genes. Within the framework of our 
results, this means that, as the genome size increases, the networks get more 
stable. However, with increased number of genes, multicellularity becomes 
common. Including interactions between cells should make the overall system 
less stable. Indeed, when investigating this issue by simulations of a 
simplified tissue model, the stability decreases with interconnectivity 
between the cells, $\kappa$, as can be seen from Fig.~4. 

We predict how the average number of active genes increases with 
$\gamma$. This may not be easy to verify, given that such a number will 
depend upon experimental conditions. {{}{}}{{}}{{{}}}{{{{}{}}}}
It should be pointed out that the order of magnitude of active genes is set 
by the rule generator in Eq.~\ref{eq: nestcan}, which is derived from
from fitting to the observed rules in \cite{harris}, many of which originate from 
{\it Drosophila melanogaster}. The fitted parameter $\alpha$ sets the
scale of the fraction of active genes, with high $\alpha$ corresponding to
low gene activity and vice versa. The qualitative behavior is, however,
rather insensitive to the value of $\alpha$.

In summary, we have designed and analyzed a class of Boolean genetic network 
models consistent with observed interaction rules. The emerging
ensemble properties do not only exhibit remarkable stability 
for the dynamics, but also several generic properties that make 
predictions, such as how stability varies with genome size, and how
the number of active genes depends on the in-degree distribution. Since
the single-cell calculations are performed analytically, the results are
transparent in terms of understanding the underlying dynamics.

\newpage

\section*{Appendix A -- Robustness} 

The stability measure, $r$, expresses the average number of perturbed
nodes one time step after perturbing one node, given that the network
has reached equilibrium in a mean field sense. Both the mean field
equilibrium distribution (of `true' and `false') and $r$ are closely
related to attractors in the true, non-mean field dynamics. See {\it
Appendix B}.

Let $W(w)$ denote the probability that a randomly chosen rule will
output `true', given that the input values are randomly and
independently chosen with probability $w$ to be `true'.
Let $W_K(w)$ denote $W(w)$ when the selected rule has $K$ inputs.
Then, the equilibrium probability for an $N$-node network, $\weq$, satisfies
\beq
  \weq = W(\weq) = \sum_{K=1}^N p^N_K W_K(\weq)~.
  \label{eq: weq}
\eeq
Eq.~\ref{eq: weq} can be solved numerically for nested canalyzing
rules, taking advantage of the
fast (exponential) convergence of $W_K(w)$ as $K \rightarrow \infty$
and using standard (integration-based) methods to calculate the sum in
the case that $N \rightarrow \infty$. Note that $\weq$ is only
referring to the mean field equilibrium distribution, which is
essentially the same as, but not identical to, the
distribution of `true' and `false' after many time steps in a
simulation.

For nested canalyzing rules, $W_K(w)$ is given by
\beq
\begin{split}
  W_K(w) = \sum_{k=1}^K\,&P_1^{\srm{nc}}\cdots P^{\srm{nc}}_{k-1}
           P_k^{\srm{can}}P(O_k = \TRUE)\\
         +\,&P_1^{\srm{nc}}\cdots P_K^{\srm{nc}}P(\Odef = \TRUE)
\end{split}
\eeq
where $P_k^{\srm{can}}=P(i_k = I_k)$ and $P_k^{\srm{nc}}=P(i_k \ne I_k)$.
The input values, $i_1,\ldots,i_K$, are `true' with probability $w$,
while the corresponding probabilities for $I_1,\ldots,I_K$ and
$O_1,\ldots,O_K$ are given by Eq.~\ref{eq: nestcan}.

Let $r(R)$ denote the probability that the rule $R$ is dependent on a
randomly picked input, given that the other inputs are independently
set to `true' with probability $\weq$. We can express $r$ for
a specific $N$-node network with rules $R_1,\ldots,R_N$ as
\beq
  r = \sum_{i=1}^N r(R_i)\frac{K(R_i)}N = \frac1N\sum_{i=1}^N S(R_i)
\eeq
where $K(R_i)$ is the number of inputs to $R_i$. We have defined
$S(R) = K(R)r(R)$, so that $r$ is the average of $S(R)$
over all rules in the network. This is also valid for a distribution
of networks, meaning that
\beq
  r_N = \sum_{K=1}^N p^N_K S_K
  \label{eq: rN}
\eeq
for $N$-node networks, where $S_K$ is the average of $S(R)$ when
random $K$-input rules are chosen.

Eq.~\ref{eq: rN} is exact, given that the state of the network is
randomly picked from the mean field equilibrium distribution of `true'
and `false'. Since the derivations are completely independent of the
specific network architecture, this result holds for any procedure
to generate architectures, as long as the average fraction of nodes with
$K$ inputs are given by $p_K^N$ (over many network realizations).

For nested canalyzing rules, we can calculate $S_K$ as a sum of
probabilities. If the rule is canalyzing on input $k$, and changing
$i_k$ makes the rule canalyze on input $j$, there is some probability
that the output value changes. The case that the rule falls back to
$\Odef$ corresponds to the last term in the square brackets.
\beq
\begin{split}
  S_K = \sum_{k=1}^{K-1}P_1^{\srm{nc}}\cdots P_{k-1}^{\srm{nc}}
    \biggl[\sum_{\,j=k+1\!}^K\!&P_{k+1}^{\srm{nc}}\cdots
     P_{j-1}^{\srm{nc}}P_j^{\srm{can}}P(O_k \ne O_j)\\
           +\,&P_{k+1}^{\srm{nc}}\cdots P_K^{\srm{nc}}
    P(O_k \ne \Odef)\biggr]
\end{split}
\eeq
where $P_k^{\srm{can}}=P(i_k = I_k)$ and $P_k^{\srm{nc}}=P(i_k \ne I_k)$.

Let $v$ and $V(v, w)$ denote the fraction of input and
output values, respectively, that are toggled (from `false' to
`true' or vice versa) when going one step forward in time,
given that the fraction `true' input values is $w$ both before and
after the input is toggled. Then, $V(0, w) = 0$,
since constant input renders constant output.
A small change in $v$ will change the output with
$r$ times the change $v$. This means
that $\partial_v V(v, \weq[]) \leq r$,
where the inequality comes from the possibility that new
changes undo previous ones, as $v$ is increasing.
Combining these relations yields
\beq
  V(v, \weq[]) \leq rv
  \label{eq: v-ineq}~,
\eeq
which means that, for $r < 1$, $V(v, \weq[]) \leq v$ with equality
if and only if $v=0$. Hence, frozen states, where the fraction of
`true' nodes is $\weq[]$, are the only solutions to the mean-field
dynamics, given that $r < 1$, which is true for the nested
canalyzing rule distribution.

Note that $v$ can be seen as the distance, {\it i.e.}, fraction of differing
states, between two separate time series. Then, the mapping $v \mapsto
V(v,\weq[])$ gives the evolution of the distance during one time
step. Similar calculations have been carried out in
{\it e.g.}, \cite{aldana2, derrida2}, yielding results consistent with
Eq.~\ref{eq: v-ineq}.


\section*{Appendix B -- Attractors}

To calculate the average number of attractors, we use the same
approach as in \cite{prl}. This approach means that we first transform
the problem of finding an $L$-cycle to a fixed point problem, and then find
a mathematical expression for the average number of solutions to that
problem.

Assume that a Boolean network performs an $L$-cycle. Then, each node
performs one of $2^L$ series of output values. We call these {\it
$L$-cycle series}. Consider what a rule does when it is subjected to
such $L$-cycle series on the inputs. It performs some Boolean
operation, but it also delays the output, giving a one step difference
in phase for the output $L$-cycle series. If we view each $L$-cycle
series as a state, an $L$-cycle turns into a fixed point (in this
enlarged state space). $L\CL$ is then the average number of input
states (choices of $L$-cycle series), for the whole network, such that
the output is the same as the input.

Let $\Q$ denote a specific choice of $L$-cycle series for the network,
and let $\QNL$ be the set of all $\Q$ such that $\Q$ is a proper
$L$-cycle. A proper $L$-cycle has no period shorter than $L$.
The average number of $L$-cycles is then given by
\beq
  \CL = \frac1L\sum_{\Q \in \QNL}P_L^N(\Q)
\eeq
where $P_L^N(\Q)$ denotes the probability that the output of the network
is the same as the input $\Q$.

Since the inputs to each $K$-input rule are chosen from a flat
distribution over all nodes, $P_L^N(\Q)$ is invariant with respect to permutations
of the nodes.  Let $\nb = (n_0, \ldots, n_{m-1})$ denote the number
of nodes expressing each of the $m=2^L$ series. For $n_i$, we refer
to $i$ as the {\it index} of the considered $L$-cycle series. For convenience,
let the constant series of `false' and `true' have indices 0 and 1,
respectively. Then,
\beq
  \CL = \frac1L\sum_{\nb \in \PNL}\binom{N}{\nb}P_L^N(\Q)
  \label{eq: CL}
\eeq
where $\binom{N}{\nb}$ denotes the multinomial
$N!/(n_0!\cdots n_{m-1}!)$ and $\PNL$ is the set of partitions
$\nb$ of $N$ such that $\Q \in \QNL$. That is, $\nb$ represents
a proper $L$-cycle.

Let $A_L^i(\x)$ denote the probability that a randomly selected rule
will output $L$-cycle series $i$, given that the input
series are selected from the distribution $\x = (x_0,\ldots,x_{m-1})$.
Since a node may not be used more than once as an input to a specific
rule, $A_L^i(\x)$ should also depend on $N$. However, the difference
between allowing or not allowing coinciding inputs vanishes as $N$
goes to infinity, since the output is effectively determined by
relatively few inputs for nested canalyzing rules.

Since these calculations aim to reveal the asymptotic behavior as
$N\rightarrow\infty$, we allow for coinciding inputs in the following.
Then, we get
\beq P_L^N(\Q) = \prod_{\substack{0\leq i<m\\ n_i\neq 0}}\!
  \bigl[A_L^i(\nb/N)\bigr]^{n_i}~.
  \label{eq: P(Q)}
\eeq

By combining Eqs.\ \ref{eq: CL} and \ref{eq: P(Q)} and applying
Stirling's formula to $\binom{N}{\nb}$, we get
\beq
  \CL \approx \frac1L\sum_{\nb \in \PNL} \frac{\sqrt{2\pi N}}
      {\dst \prod_{\substack{0\leq i<m\\ n_i\neq 0}}\!\!
	\sqrt{2\pi n_i}}e^{N f_L(\nb/N)}
  \label{eq: CL-approx}
\eeq
where
\beq
  f_L(\x) = \sum_{\substack{0\leq i<m\\ x_i\neq 0}}\!
          x_i \ln\biggl[\frac1{x_i}A_L^i(\x)\biggr]~.
  \label{eq: fL}
\eeq
See {\it Supporting Text} regarding the use of Stirling's formula.

Equation \ref{eq: fL} can be seen as an average
$\bigl<\ln\bigl[\frac1{x_i}A_L^i(\x)\bigr]\bigr>_i$
with weights $x_0,\ldots,x_{m-1}$. Hence, the concavity of
$x \mapsto \ln x$ yields
\beq
\begin{split}
  f_L(\x) &= \biggl\langle\ln\biggl[\frac1{x_i}A_L^i(\x)\biggr]\biggr\rangle_
                  {\!\!i}
          \\&\leq \ln\,\biggl\langle\frac1{x_i}A_L^i(\x)\biggr\rangle_
                  {\!\!i}
          = \ln \!\sum_{\substack{0\leq i<m\\ x_i\neq 0}}
                        \! A_L^i(\x)
          \leq 0 
\end{split}
\eeq
with equality if and only if
\beq
  \x = \A_L(\x)
  \label{eq: mean-field}
\eeq
where $\A_L(\x) = \bigl(A_L^0(\x),\ldots,A_L^{m-1}(\x)\bigr)$. Eq.~\ref{eq:
mean-field} can be seen as a criterion for mean field equilibrium in
the distribution of $L$-cycle series. This makes it possible to
connect quantities observed in mean field calculations to the full
non-mean field dynamics.

Since $Nf_L(\x)$ occurs in the exponent in Eq.~\ref{eq: CL-approx},
the relevant contributions to the sum in Eq.~\ref{eq: CL-approx} must
come from surroundings to points where Eq.~\ref{eq: mean-field} is
satisfied as $N \rightarrow \infty$. ($f_L(\x)$ is not a continuous
function, but obeys the weaker relation ``For all $\xb$, in the
definition set of $f_L$, and $\epsilon > 0$, there is a $\delta>0$
such that $f_L(\xb')<f_L(\xb)+\epsilon$ holds for all $\xb'$
satisfying $|\xb' - \xb| < \delta$ in the definition set of $f_L$.''
which is a sufficient criterion in this case.) Eq.~\ref{eq:
CL-approx} provides an upper bound to $\CL$, since the approximation
of $\binom{N}{\nb}$ is an overestimation, see {\it Supporting Text}.
Hence, the relevant regions in the exact sum,
Eq.~\ref{eq: CL}, surround solutions to Eq.~\ref{eq: mean-field}.

Given that Eq.~\ref{eq: mean-field} holds, the fraction of `true'
values, $w(\x)$, and the fraction of togglings $v(\x)$, in the
distribution of $L$-cycle series, should be consistent with the mean
field dynamics, meaning that $w(\x)=\weq[]$ and $v(\x)=0$. (See {\it
Appendix A}.) This means that a typical attractor, for large $N$, has
only a small fraction (which approaches 0 for $N \rightarrow \infty$)
of active (non-constant) nodes.

For large $N$, we want to investigate the number of attractors for
certain numbers of active nodes. Hence, we divide the summation in
Eq.~\ref{eq: CL} into constant and non-constant patterns. In order
to give the sum a form that can be split in a convenient way, we
introduce the quantity $\OL$, which we define as the average number of
states that are part of a cycle with a period that divides $L$.
See {\it Supporting Text} on how to express $\CL$ in terms of $\OL$.
As $N\rightarrow\infty$, the summation over constant patterns has
a limit that yields
\beq
  \OLinf = \frac1{1-\dr}\sum_{\nbh\in\nat^{m-2}}
            \prod_{\substack{2\leq i<m\\ n_i\neq 0}}
            e^{-rn_i}
            \frac{(\nbh\cdot\nabla A_L^i)^{n_i}}{n_i!}~.
  \label{eq: OLinf0}
\eeq
where $\nbh=(n_2,\ldots,n_{m-1})$, and $\nat$ denotes the
set of non-negative integers. See {\it Supporting Text}.

The elements $\partial_jA_L^i$ of the gradient $\nabla A_L^i$ are
nonzero only if the $L$-cycle series $i$ can be the output of a node
which has only one non-constant input retrieving the $L$-cycle series
$j$. This is true if the series $j$ is the series $i$, or the inverse
of $i$, rotated one step backwards in time. Let $\phiC(i)$ and
$\phiI(i)$ denote those values of $j$, respectively.
Then,
\beq
  \nbh\cdot\nabla A_L^i = \rC n_{\phiC(i)}+\rI n_{\phiI(i)}~.
  \label{eq: OLinf1g}
\eeq
%
%
Equation \ref{eq: OLinf1g} yields that the sum in Eq.~\ref{eq:
OLinf0} factorizes into subspaces, spanned by sets of $L$-cycle series
indices of the type $\bigl\lbrace
i,\phiC(i),\phiI(i),\phiC\circ\phiC(i),\phiC\circ\phiI(i),\ldots\bigr\rbrace$
containing all possible results of repeatedly applying $\phiC$ and
$\phiI$ to $i$. We call those sets {\it invariant sets of $L$-cycle
series}, which is the same as {\it invariant sets of $L$-cycle
patterns} in \cite{prl}, but formulated with respect to $L$-cycle
series instead of $L$-cycle patterns. Let
$\rho_L^0,\ldots,\rho_L^{H_L-1}$ denote the invariant sets of
$L$-cycle series, where $H_L$ is the number of such sets. For
convenience, let $\rho_L^0$ be the invariant set $\lbrace0,1\rbrace$.

Consider an invariant set of $L$-cycle series, $\rho_L^h$. Let $\ell$ be
the {\it length} of $\rho_L^h$, meaning that $\ell$ is the lowest number
such that, for $i \in \rho_L^h$, $(\phiC)^\ell(i)$ is either
$i$ or the index of series $i$ inverted. If $(\phiC)^\ell(i) =
i$, we say that the {\it parity} of $\rho_L^h$ is positive. Otherwise
the parity is negative. The structure of an invariant set of $L$-cycle
series is fully determined by its length and its parity. Such a set can be
enumerated on the form $\bigl\lbrace\phiC(i),\ldots,(\phiC)^\ell(i),
\phiI(i),\ldots,\phiI\circ(\phiC)^{\ell-1}(i)\bigr\rbrace$ and
$(\phiC)^\ell(i) = i$ for positive parity while
$\phiI\circ(\phiC)^{\ell-1}(i) = i$ for negative parity.

Let strings of $\T$ and $\F$ denote specific $L$-cycle series. Then
$\phiC(\F\F\F\T) = \F\F\T\F$ and $\phiI(\F\F\F\T) = \T\T\F\T$.
Examples of invariant sets of $4$-cycle series are $\lbrace\F\F\F\T$,
$\F\F\T\F$, $\F\T\F\F$, $\T\F\F\F$, $\T\T\T\F$, $\T\T\F\T$,
$\T\F\T\T$, $\F\T\T\T\rbrace$ and $\lbrace\F\T\F\T$,
$\T\F\T\F\rbrace$. The first example has length $4$ and positive
parity, while the second has length $1$ and negative parity.

Let $\nu^+_1,\nu^-_1,\ldots,\nu^+_\ell,\nu^-_\ell$ denote the numbers,
$n_i$, of occurrences of $L$-cycle series belonging to $\rho_L^h$, in
such a way that $\nu^\pm_{\ip}\equiv
n_i\Leftrightarrow\nu^\pm_{\ip-1}\equiv n_{\phiC(i)}$ and
$\nu^\pm_{\ip}\equiv n_i\Leftrightarrow\nu^\mp_{\ip-1}\equiv
n_{\phiI(i)}$. For convenience, we introduce $\nu^\pm_0$ as a renaming
of $\nu^\pm_\ell$. There are two ways that $\nu^\pm_0$ can be connected
to $\nu^\pm_\ell$: either $\nu^\pm_0 \equiv \nu^\pm_\ell$ (positive
parity) or $\nu^\pm_0 \equiv \nu^\mp_\ell$ (negative parity).

Each invariant set of $L$-cycle series, $\rho_L^h$, contributes to
Eq.~\ref{eq: OLinf0} with a factor
\beq
  g(\rho_L^h) \equiv g_\ell^\pm = \!\!\!
      \sum_{\nu^+\!\!,\nu^-\in\nat^\ell}\,
      \prod_{\ip=1}^\ell
      G_{\nu^+_\ip\nu^-_\ip}^{\nu^+_{\ip-1}\nu^-_{\ip-1}}
  \label{eq: gl0}
\eeq
where
\begin{align}
  \label{eq: G}
  G_{\nu^+_\ip\nu^-_\ip}^{\nu^+_{\ip-1}\nu^-_{\ip-1}}
   &= 
    \exp(-\tilde{\nu}^+_\ip)\frac{(\tilde{\nu}^+_\ip)^{\nu^+_\ip}}
      {\nu^+_\ip!}
    \exp(-\tilde{\nu}^-_\ip)\frac{(\tilde{\nu}^-_\ip)^{\nu^-_\ip}}
      {\nu^-_\ip!}\\
  \tilde{\nu}^\pm_\ip &\equiv \rC\nu^\pm_{\ip-1}+\rI\nu^\mp_{\ip-1}
   ~,
\end{align}
and $\nu^\pm_0 \equiv \nu^\pm_\ell$ for $g_L^+$ while $\nu^\pm_0 \equiv
\nu^\mp_\ell$ for $g_L^-$. Eq.~\ref{eq: G} is interpreted with the
convention that $0^0 = 1$ to handle the case where $\nu^+_\ip$ or
$\nu^-_\ip$ $=0$. 

Although the right hand side in Eq.~\ref{eq: gl0} looks nasty,
it can be calculated yielding the expression
\beq
    g_\ell^\pm = \frac1{1-r^\ell}\frac1{1\mp(\dr)^\ell}~.
  \label{eq: gl-final}
\eeq
See {\it Supporting Text}.

Now, we can write Eq.~\ref{eq: OLinf0} as
\beq
  \OLinf = \frac1{1-\dr}
            \prod_{h=1}^{H_L-1}g(\rho_L^h)
  \label{eq: OLinf2}
\eeq
where $g(\rho_L^h)$ is calculated according to Eq.~\ref{eq:
gl-final}. The period, $\ell$, and the parity, $+$ or $-$, of a given
invariant set of $L$-cycle series can be extracted by enumerating all
$L$-cycle series. This provides a method to calculate $\OLinf$ for
small $L$. See {\it Supporting Text} on how to calculate
$\OLinf$ in an efficient way.

\subsection*{Acknowledgments}

CT acknowledges the support from the Swedish National Research School in 
Genomics and Bioinformatics. CP is affiliated with the Lund Strategic Center 
for Stem Cell Biology and Cell Therapy and SK is affiliated with the Santa Fe 
Institute.

{}
\newpage

\parindent 0pt
\parskip 3ex

{\Large\bf Figure captions}

{\bf Fig.~1.}\quad
Examples of $N=20$ networks with $\gamma = 1$ {\bf(a)}, $\gamma = 2$ {\bf(b)} 
and $\gamma = 3$ {\bf(c)}.

{\bf Fig.~2.}\quad 
Robustness, $r$, and the probability of a node being `true', $w$,
at the equilibrium distribution of Boolean values, as functions of the
exponent $\gamma$ in the in-degree distribution
$p_K$ (Eq.~\ref{p_in}) for $N=20$ (dotted), 100 (dashed)
and $\infty$ (solid).

{\bf Fig.~3.}\quad 
The number of attractors as function of the exponent $\gamma$
in the in-degree distribution $p_K$ (Eq.~\ref{p_in}) for $N=20$ (thick lines)
and $N \rightarrow \infty$ (thin lines). The curves show the cumulative
number of attractors of length $L$ for $L=1$ (solid), $L\leq2$ (dashed) and
$L\leq\infty$ (dotted).
The values for $N\rightarrow\infty$ were calculated analytically,
whereas the values for $N=20$ are taken from full enumerations of
the state space for at least 5000 networks, with more networks at
higher $\gamma$.

{\bf Fig.~4.}\quad 
The average time evolution of perturbed fixed points in $5 \times 5$ cell
tissues with periodic boundary conditions and $N=50$ nodes (over many
network realizations). Simulations
from random initial states in generated networks were used to locate
fixed points, which were perturbed by toggling the value of a single
node. The mean distances to the unperturbed fixed point,
$\langle H(t) \rangle$, as given by 20 subsequent simulation steps,
is shown for $\gamma = 2$ \textbf{(a)} and
$\gamma = 3$ \textbf{(b)}, for three different degrees of cell connectivity:
$\kappa$ = 0 (solid), 0.05 (dashed) and 0.1 (dotted).

\newpage

\begin{center}
\rotatebox{0}{
\hspace{-15mm}
\epsfig{figure=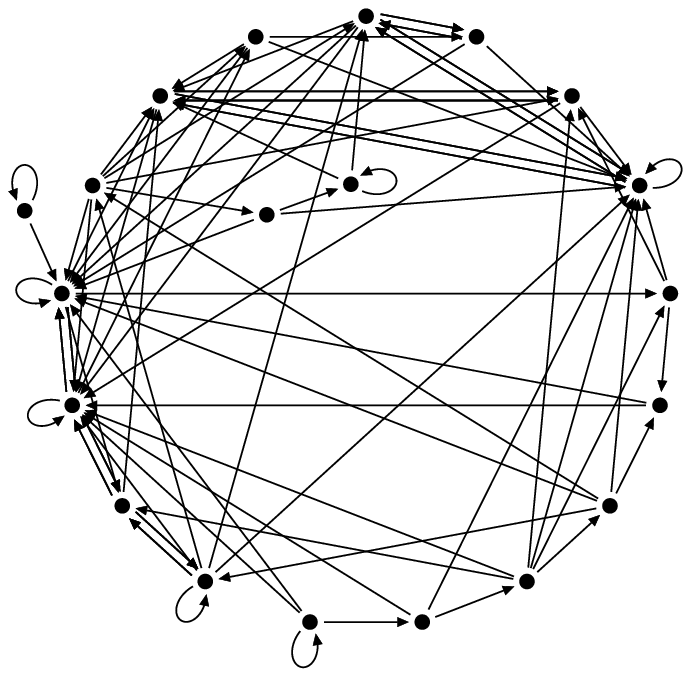,width=9cm}
\epsfig{figure=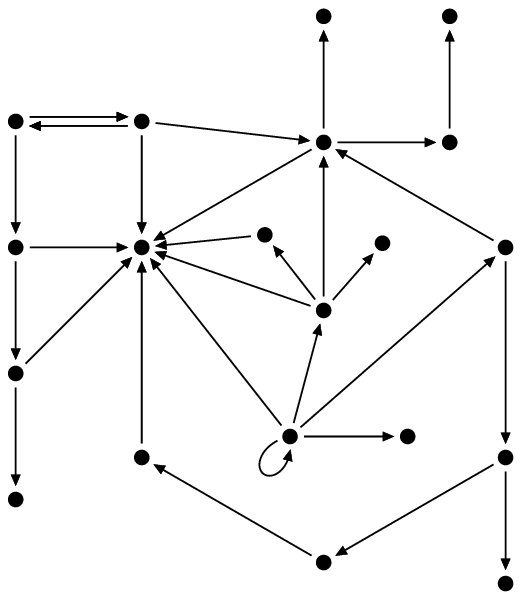,width=9cm}
}
\end{center}

\begin{center}
{\bf Fig.~1a} \hspace{80mm} {\bf Fig.~1b}
\end{center}

\begin{center}
\rotatebox{0}{
\epsfig{figure=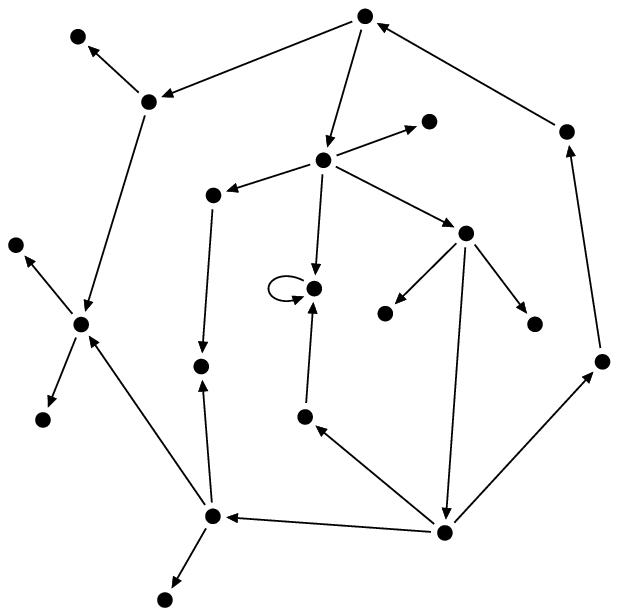,width=9cm}
}
\end{center}

\begin{center}
{\bf Fig.~1c}
\end{center}

\newpage

\begin{center}
\epsfig{figure=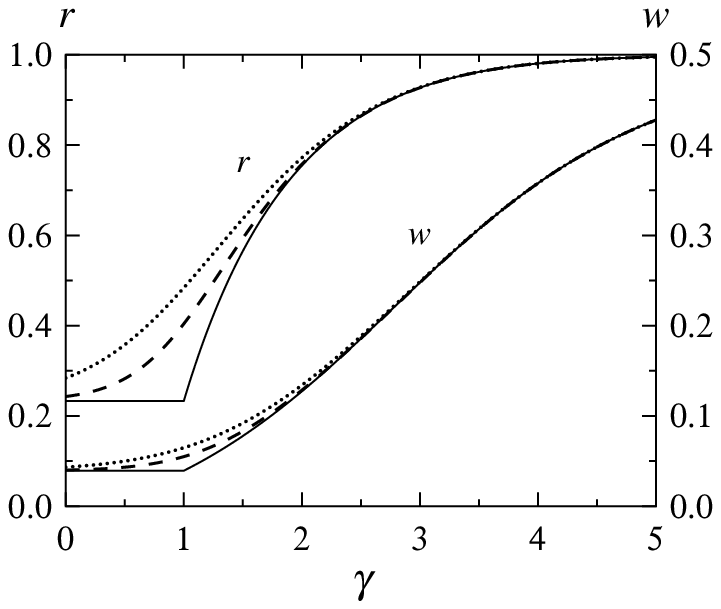,width=10cm}
\end{center}

\begin{center}
{\bf Fig.~2}
\end{center}

\begin{center}
\epsfig{figure=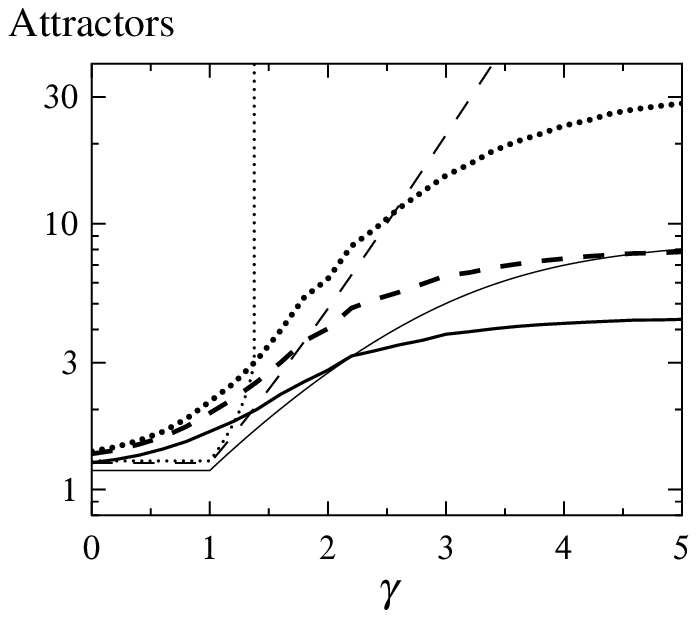,width=10cm}
\end{center}

\begin{center}
{\bf Fig.~3}
\end{center}

\newpage

\begin{center}
\epsfig{figure=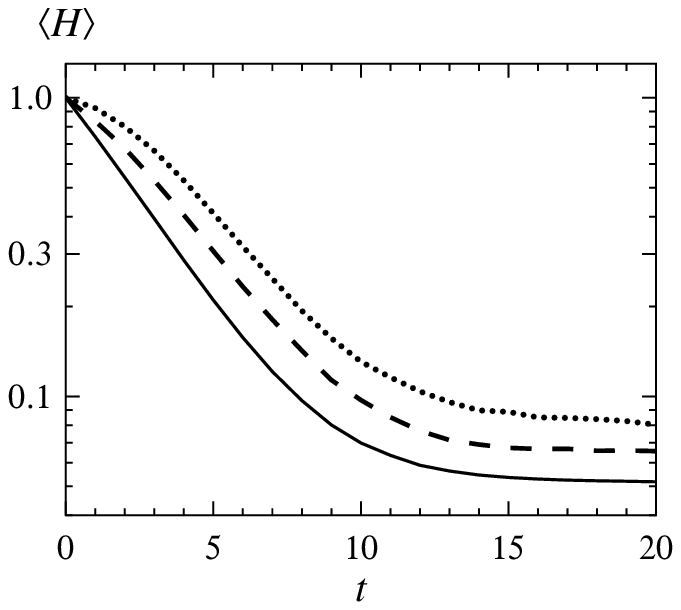,width=10cm}
\end{center}

\begin{center}
{\bf Fig.~4a}
\end{center}

\begin{center}
\epsfig{figure=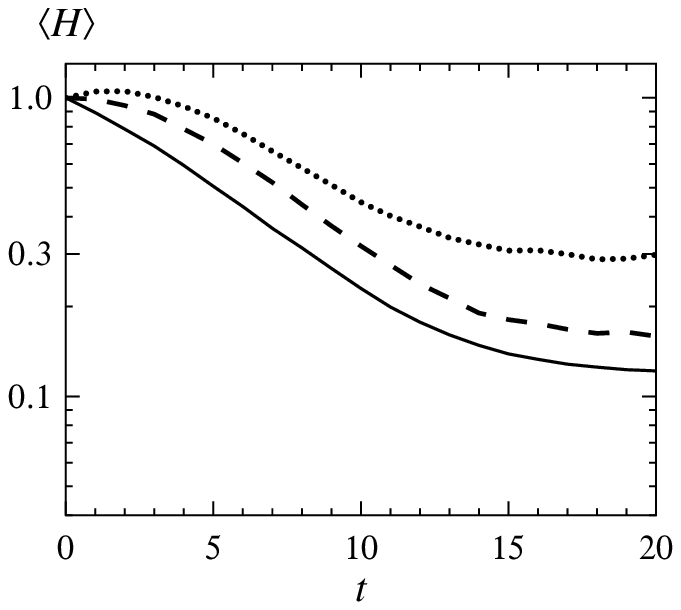,width=10cm}
\end{center}

\begin{center}
{\bf Fig.~4b}
\end{center}

\end{document}